\documentclass[twocolumn,showpacs,preprintnumbers,amsmath,amssymb,prl,superscriptaddress]{revtex4-1}
\newcommand{\BE}{\textsc{be}}

\newcommand{\FSR}{\textsc{fsr}}
\newcommand{\HEP}{\textsc{hep}}
\newcommand{\HFI}{\textsc{hfi}}
\newcommand{\PBC}{\textsc{pbc}}
\newcommand{\SST}{\textsc{sst}}
\newcommand{\scylla}{Scylla}
\newcommand{\charybdis}{Charybdis}
\newcommand{\scA}{\ensuremath{\mathscr A}}
\newcommand{\scD}{\ensuremath{\mathscr D}}

\newcommand{\scL}{\ensuremath{\mathscr L}}
\newcommand{\scO}{\ensuremath{\mathcal O}}
\newcommand{\scS}{\ensuremath{\mathcal S}}
\newcommand{\sD}{\ensuremath{\EuScript D}}
\newcommand{\specterm}[3]{\ensuremath{^#1\!#2_{#3}}}
\newcommand{\ssz}{\specterm1S0}%
\newcommand{\spo}{\specterm1P1}%
\newcommand{\tdo}{\specterm3D1}%
\newcommand{\tso}{\specterm3S1}%
\newcommand{\tpx}{\specterm3P{}}%
\newcommand{\BFpm}{\raisebox{.1em}{\color{blue}$+$}\hspace{-.775em}\raisebox{-.225em}{\color{red}$-$}}
\newcommand{\BFpmss}{{\raisebox{.12em}{\color{blue}$\scriptstyle+$}\hspace{-.6177em}\raisebox{-.125em}{\color{red}$\scriptstyle{}-$}}}
\DeclareRobustCommand{\LCBFpmss}{\raisebox{.12em}{\color{blue}$\scriptstyle+$}\hspace{-.6177em}\raisebox{-.12em}{\color{red}$\scriptstyle{}-$}}

\newcommand{\boseplus}{{\ensuremath{\color{blue}+}}}
\newcommand{\ferminus}{{\ensuremath{\color{red}-}}}
\newcommand{\oneover}[1]{\frac{1}{#1}}
\newcommand{\proj}[1]{\ket{#1}\bra{#1}}
\newcommand{\bra}[1]{\ensuremath{\left\langle#1\right\vert}}
\newcommand{\ket}[1]{\ensuremath{\left\vert#1\right\rangle}}
\newcommand{\abs}[1]{\left|#1\right|}
\newcommand{\prn}[1]{\left(#1\right)}
\newcommand{\brc}[1]{\left\{#1\right\}}
\newcommand{\nm}[3]{\mbox{\if#11{$10^{#2}$}\else{\ifnum#2=0{$#1$}\else{$#1\times10^{#2}$}\fi}\fi{ }#3}}
\newcommand{\ts}[1]{_{\mbox{\scriptsize{}#1}}}
\newcommand{\tu}[1]{^{\mbox{\scriptsize{}#1}}}
\newcommand{\figref}[1]{\textsc{Fig.}~\ref{#1}}
\newcommand{\eqnref}[1]{\textsc{Eq.}~\eqref{#1}}
\usepackage{graphicx}
\usepackage{color}
\usepackage{euscript}
\usepackage{mathrsfs}
\usepackage{bm}
\definecolor{darkgreen}{rgb}{.12, .5, 0}
\begin{document}
\title{Spectroscopic test of Bose-Einstein statistics for photons}%
\newcommand{\berkeley}{{Department of Physics, University of California, Berkeley, California 94720-7300, USA}}
\author{D. English}
\email{denglish@berkeley.edu}
\affiliation{\berkeley}%
\author{V. V. Yashchuk}
\affiliation{Advanced Light Source, Lawrence Berkeley National Laboratory, Berkeley, California 94720, USA}%
\author{D. Budker}
\affiliation{\berkeley}%
\affiliation{Nuclear Sciences Division, Lawrence Berkeley National Laboratory, Berkeley, California 94720, USA}%

\date{\today}
%
%
\begin{abstract}
Using Bose-Einstein-statistics-forbidden two-photon excitation in atomic barium, we have limited the rate of statistics-violating transitions, as a fraction $\nu$ of an equivalent statistics-allowed transition rate, to $\nu<4.0\times10^{-11}$ at the 90\% confidence level. This is an improvement of more than three orders of magnitude over the best previous result. 
Additionally, hyperfine-interaction enabling of the forbidden transition has been observed, to our knowledge, for the first time.
\end{abstract}

\pacs{42.50.Xa, 82.50.Pt}

\maketitle
%
%
%
The Spin-Statistics Theorem (\SST) is proved in the framework of relativistic field theory using the assumptions of causality and Lorentz invariance in 3+1 spacetime dimensions.  This letter describes an experimental test of Bose-Einstein (\BE) statistics and, consequently, the \SST\ as it applies to photons interacting with atoms.  As such, the experiment is a test of the assumptions,
%
both explicit and subtle \cite{Wich01}, of the \SST\ and the assumptions made in the quantization of the photon field.

\begin{figure}
\includegraphics[width=2.30in]{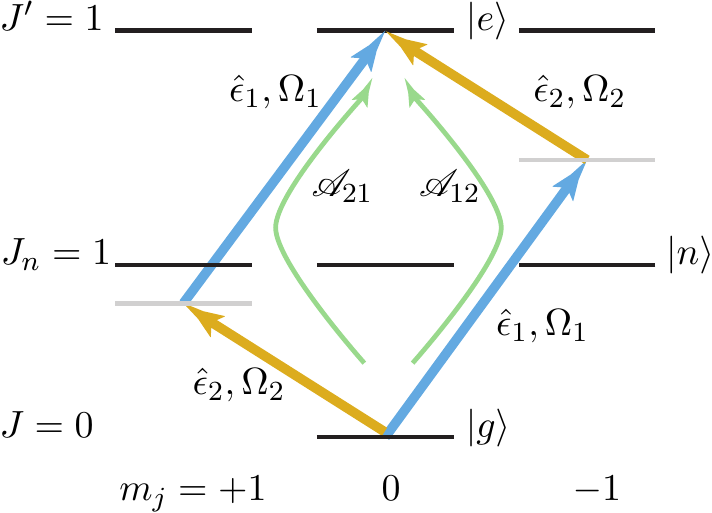}
\caption{(color online). Two amplitudes interfere in the excitation of the $m_j'=0$ sublevel in a $J=0\to J'=1$ two-photon atomic transition. Similar diagrams can be drawn for excitation to the $m_j'=\pm1$ sublevels.}\label{fig:transition}
\end{figure}

The experiment uses a selection rule \cite{Dra68,Bon84} for atomic transitions that is closely related to the Landau-Yang theorem \cite{Lan48,Yan50} in high-energy physics (\HEP).  The selection rule states that two collinear, equal-frequency, photons cannot participate in any process that would require them to be in a state of total angular momentum one.  An example in \HEP\ is that the neutral spin-one $Z_0$ boson cannot decay to two photons, $Z_0\nrightarrow\gamma\gamma$ (presently, the branching ratio for this process is limited to less than \nm{5.2}{-5}{} \cite{pdg2006}).  For atoms, the selection rule means that two collinear equal-frequency photons cannot stimulate a transition between atomic states of total angular momentum zero and one, $J=0\nrightarrow J'=1$ (\figref{fig:transition}).  The reason is the same in both cases:  The photons would have to be in a state of total angular momentum $J=1$, an exchange-\textit{anti}symmetric state, in violation of \BE\ statistics and the \SST\ (see Ref. \cite{hil00b} for a review of other examples, other tests of particle statistics, and theoretical attempts to accommodate violations of the \SST).
%
%
%
%
%
%
We illustrate the selection rule, applied to this experiment, as follows:
The rate $W$ for E1E1 two-photon excitation from an atomic ground \ket{g} to excited \ket{e} state is \cite{Gol81,Dun04}
\begin{align}\nonumber
W_\BFpmss=\frac{2\pi}{\hbar^4}& \times\abs{\sum_n {\scA^{(n)}_{12}}\BFpm\;{\scA^{(n)}_{21}} }^2\\\label{eq:twophotrate}
&\times\oneover{\pi}\frac{\Gamma/2}{\prn{\Omega_1+\Omega_2-\omega_{eg}}^2+\prn{\Gamma/2}^2}
\frac{\bar{I}_1\bar{I}_2}{4\epsilon_0^2\,c^2},
\end{align}
where
\begin{align}\label{eq:amplitude}
  \scA^{(n)}_{jk}=
  \frac{\bra{e}\hat{\epsilon}_{\color{darkgreen}k}\cdot\scD\proj{n}\hat{\epsilon}_{\color{darkgreen}j}\cdot\scD\ket{g}}
  {\omega_{ng}-\Omega_{\color{darkgreen}j}+i\Gamma_n/2}
\end{align}
is the amplitude for the ordered absorption of photons $j$ and $k$; \ket{g}, \ket{n}, and \ket{e} are the ground, intermediate, and excited states; $\Gamma_n$ and $\Gamma$ are the intermediate- and excited-state natural widths; $\hat{\epsilon}_{1,2}$, $\Omega_{1,2}$, and $\bar{I}_{1,2}$, are the polarization, frequency, and field intensity of photon 1, 2;  $\omega_{kl}\equiv\omega_k-\omega_l$ is the energy difference between states $k$ and $l$ expressed in frequency units; and $\scD$ is the electric-dipole operator.

There are two paths to the final state, hence the two amplitudes $\scA^{(n)}_{12}$ and $\scA^{(n)}_{21}$ in \eqnref{eq:twophotrate}.  Bose statistics requires that the amplitudes add with a relative `$\boseplus$' (upper) sign.  We investigate the consequences of exchange antisymmetric photon states by adding the exchange amplitudes with a relative `$\ferminus$' (lower) sign. Then, performing the sum over magnetic sublevels of the largest contributing manifold, applying the Wigner-Eckart theorem, extracting the rank-1 irreducible component of the transition operator, and assuming orthogonal light polarizations (which maximizes the rate \cite{Eng07}) give
\begin{align}\label{eq:tranrate}
W^{}_\BFpmss=\abs{f_\BFpmss}^2
\frac{\Gamma/2}{\prn{\Omega_1+\Omega_2-\omega_{eg}}^2+\prn{\Gamma/2}^2}
\frac{\sD_{en}^2\sD_{ng}^2\bar{I}_1\bar{I}_2}{3\epsilon_0^2\,c^2\hbar^4},
\end{align}
where
\newcommand{\Da}{\prn{\omega_{ng}-\Omega_1+i\Gamma_n/2}}%
\newcommand{\Db}{\prn{\omega_{ng}-\Omega_2+i\Gamma_n/2}}%
\begin{align}\label{eq:fpm:1}
    f_\boseplus=&\frac{\prn{\Omega_1-\Omega_2}/2}{\Da\Db},\\\label{eq:fpm:2}
    f_\ferminus=&\frac{\omega_{ng}-\prn{\Omega_1+\Omega_2}/2+i\Gamma_n/2}{\Da\Db},
\end{align}
and where $\sD_{ng}$ ($\sD_{en}$) is the reduced dipole matrix element of the ground- to intermediate- (intermediate- to final-) state transition.

\begin{figure}
\includegraphics[width=2.30in]{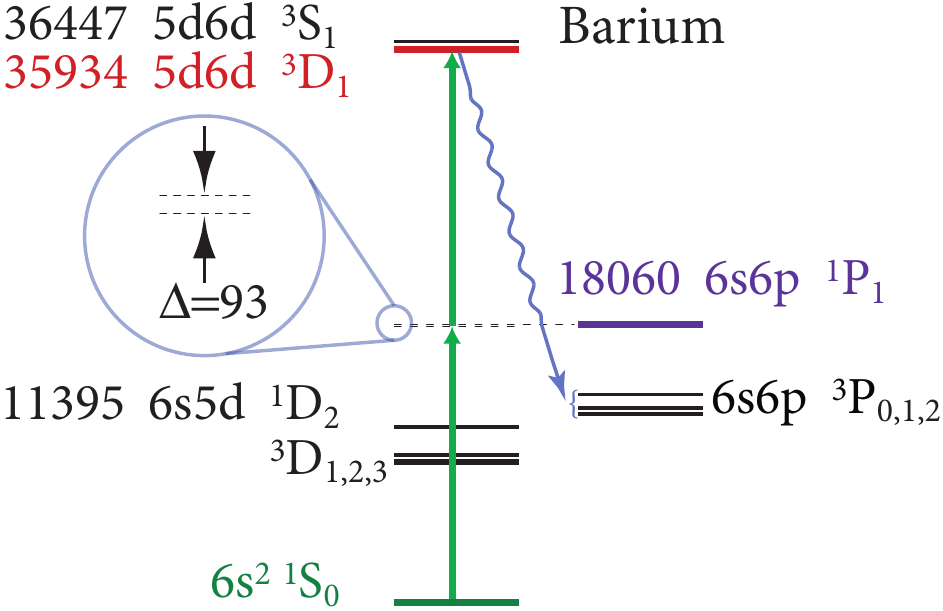}
\caption{(color online). Partial energy-level diagram of barium.  Even (odd) parity states are in the left (right) column.  Energies are in cm$^{-1}$. The close proximity of $6s6p$ \spo\ to the energy midpoint ($\Delta=18060-35934/2=\nm{93}{0}{cm$^{-1}$}$) enhances the two-photon transition rate.}\label{fig:BariumEnergyLevels}
\end{figure}

Equations \eqref{eq:fpm:1} and \eqref{eq:fpm:2} show that when the two photons have the same frequency (i.e., when $\Omega_1=\Omega_2$), then $f_{\boseplus}=0$, but $f_{\ferminus}\ne0$.
%
%
%
\begin{figure}
\includegraphics[width=240pt]{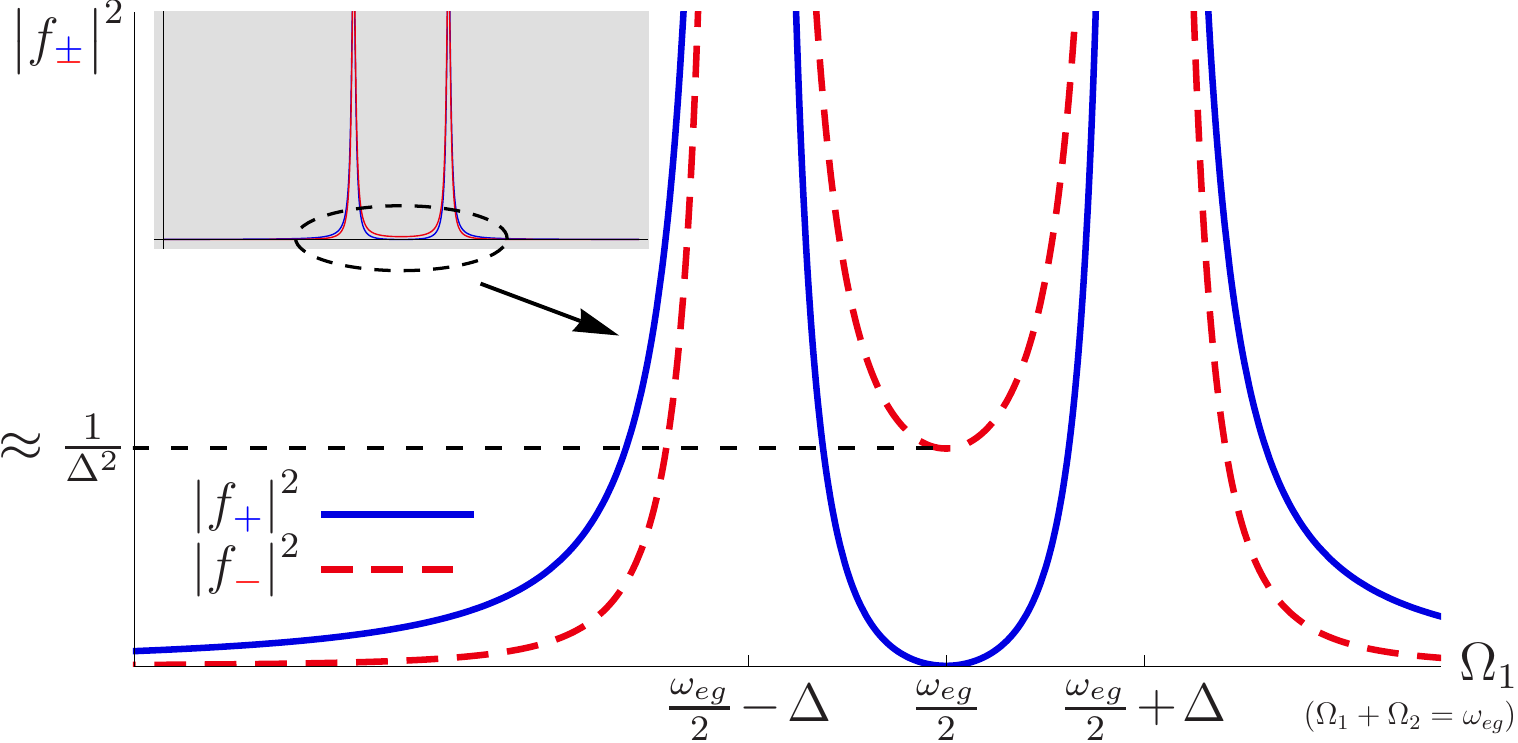}
\caption{(color online). Plot of $\abs{f_{\LCBFpmss}}^2$
on resonance ($\Omega_1+\Omega_2=\omega_{eg}$) as a function of one photon's frequency $\Omega_1$.  The two peaks correspond to the cases where the frequency of one of the photons matches the ground-to-intermediate transition.}%
\label{fig:fpmplot}
\end{figure}
Figure \ref{fig:fpmplot} shows this distinction at  $\Omega_2=\Omega_1=\omega_{eg}/2$:  for exchange-symmetric photon states, the transition rate is identically zero; for -antisymmetric states, the rate is $\propto\!\!\!1/\Delta^{2}$, where $\Delta\equiv\omega_{ng}-{\omega_{eg}}/{2}$ is the separation between the energy of the intermediate level $\omega_{ng}$,  and the halfway point $\omega_{eg}/2$.  This fact motivated the choice of atomic species in the experiment:  in barium, $\Delta$ is relatively small (\nm{93}{0}{cm$^{-1}$}), providing resonant enhancement to $W_\ferminus$ (\figref{fig:BariumEnergyLevels}).

Measurements of the transition rate made at laser frequencies near the halfway point probe the exchange symmetry of photons. It is convenient to express the rates \eqref{eq:tranrate} in the case where the two laser frequencies are symmetrically detuned by a small amount $\delta$ (where $\delta\ll\Delta$) about the energy midpoint, $\delta\equiv\abs{\omega_{eg}/2-\Omega_1}=\abs{\omega_{eg}/2-\Omega_2}$.  Assuming $\abs{\Omega_1+\Omega_2-\omega_{eg}}\ll\Gamma$, the transition rates $W_\BFpmss$ expressed as functions of $\delta$ are
\begin{align}
    W_\ferminus\prn{\delta}&=
        \frac{2\: \sD_{en}^2\sD_{ng}^2\:\bar{I}_1\bar{I}_2}{\Gamma\hbar^4\Delta^2\epsilon_0^2\,c^2}\prn{1-{\frac{\delta^2}{\Delta^2}}}^{-2},
         \\
    W_\boseplus\prn{\delta}&=
        \frac{\delta^2}{\Delta^2}\,W_\ferminus\prn{\delta}.
\end{align}
%
%
%
%
%
Because hamiltonians which treat multiple photons identically cannot mix exchange-symmetric and exchange-antisymmetric subspaces of the photon state space \cite{Ama80}, there is no interference between the two amplitudes in the total excitation rate.  Thus the measured fluorescence signal $\scS\prn{\delta}$ on resonance is
\begin{align}
  \scS\prn{\delta}&=\gamma\brc{
     W_\boseplus\prn{\delta}+\nu\,W_\ferminus\prn{\delta}
    },
\end{align}
where $\gamma$ is the product of the number of atoms and the total detection efficiency, and $\nu$ is a small dimensionless parameter that characterizes the degree of violation of \BE\ statistics.

\begin{figure}
\includegraphics[width=240pt]{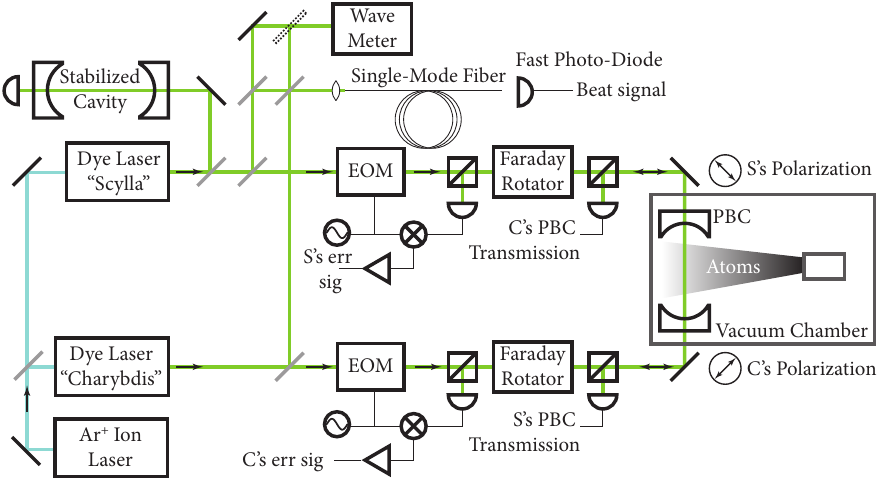}
\caption{(color online). Experimental apparatus.  An atomic beam of barium interacts with the light fields of two dye lasers ($\lambda\cong\;$557 nm) at the center of a tunable confocal optical power build-up cavity (\PBC)   of length $L=\nm{30}{0}{cm}$.  %
The lasers are coupled into the cavity to excite only symmetric transverse modes, making the effective free spectral range (\FSR) $c/2L=500$ MHz.  The cavity finesse is 340.  %
The lasers drive the $6s^2$ \ssz $\:\to5d6d$ \tdo\ two-photon transition of barium, see \figref{fig:BariumEnergyLevels}.  Fluorescence from $5d6d$ $\tdo$ to the $6s6p$ \tpx\ manifold  ($\lambda\cong\;$420 nm) is collected through a light pipe and bandpass filters (not shown), and detected by a low-noise photo-multiplier tube in the photon counting regime.  Each dye laser is locked to the \PBC\ with an electro-optic modulator (\textsc{eom}) and rf-electronics, employing the Pound-Drever-Hall method \cite{PDH}. Polarizing beam splitters and Faraday rotators isolate the lasers from each other and from their own cavity-rejected light.  The lasers are also coupled into a single-mode fiber and interference is detected with a fast photodiode.  A stabilized cavity, of \FSR$\approx$50 MHz, records transmission peaks from the ``\scylla" laser.}\label{fig:optschem}
\end{figure}

Measuring $\nu$ is achieved by first calibrating the apparatus at $\delta=\delta\ts{cal}$.  Expanding $\scS\prn{\delta}$ about $\delta=0$, for $\nu\ll\delta^2\ts{cal}/\Delta^2\ll1$, the calibration signal is
\begin{align}\label{eq:calS}
  \scS\ts{cal}=\scS\prn{\delta\ts{cal}}&=\gamma W_\ferminus\prn{0}\,\frac{\delta\ts{cal}^2}{\Delta^2}+\scO\prn{\frac{\delta\ts{cal}^4}{\Delta^4}}.
\end{align}
Measuring at the halfway point ($\delta=0$), the signal is $\scS\ts{lim}=\scS\prn{0}=\gamma\nu\,W_\ferminus\prn{0}$, so that
\begin{align}\label{eq:limit}%
    \nu&=\frac{\scS\ts{lim}}{\scS\ts{cal}}\;\frac{\delta\ts{cal}^2}{\Delta^2}.
\end{align}
%
%
%
%
%
The previous version of the experiment \cite{DeM99} measured $\nu<10^{-7}$ at the 90\% confidence level, using barium in a vapor cell and pulsed lasers to drive the $6s^2$ $\ssz\to5d6d$ $\tso$ two-photon transition.  As described below, the experiment's sensitivity was limited by the relatively large bandwidth of the lasers ($\approx3$ GHz). This letter describes a new experiment that avoids these limitations by using continuous-wave lasers, an in-vacuum power build-up cavity (\PBC), and an atomic beam (\figref{fig:optschem}).
%
%
%
%
%
\begin{figure}
\includegraphics[width=240pt]{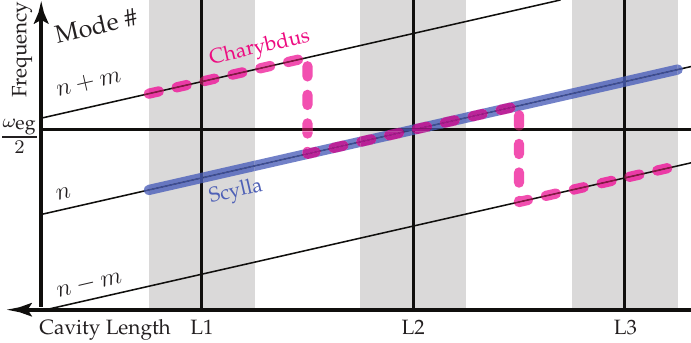}
\caption{(color online). Laser tuning path during run.  The lasers are tuned in concert with the \PBC, either separated by $m$=4 longitudinal cavity modes, or in the same mode.}\label{fig:Tuning_Scheme}
\end{figure}
%
%
%

\begin{figure}%
\includegraphics[width=240pt]{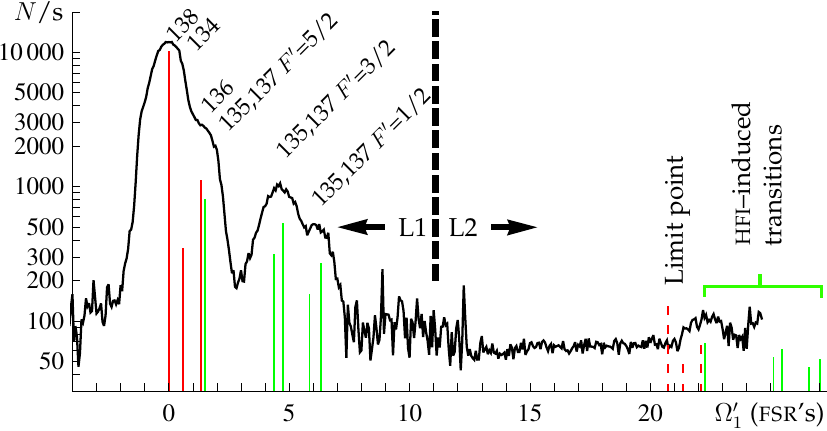}
\caption{(color online).  Processed data from one run. Fluorescence, in photons per second, from $5d6d$ \tdo\ to the $6s6p$ \tpx\ manifold.  $\Omega_1'\equiv\Omega_1-\omega^{(138)}_{eg}/2+\delta\ts{cal}$, measured in reference cavity \FSR's ($\approx$50 MHz).  In the region L1 (\figref{fig:Tuning_Scheme}), $-4\le\Omega_1'<11$ , $\Omega_2-\Omega_1=2\pi\times\nm{2}{0}{GHz}$,  and a calibration spectrum is acquired containing both even and odd isotopes.  Solid lines mark their approximate locations, known from a separate measurement to be presented elsewhere.  Light-shifts are responsible for the broad line shapes.  In the region L2, $11<\Omega_1'\le25$, $\Omega_2-\Omega_1\approx0$, many traversals have been combined for improved signal-to-noise ratio.  The spectrum of even isotopes (shown as dashed lines) is suppressed below detection by \BE-statistics.   In the spectrum of odd isotopes however, the hyperfine interaction enables the transition. This is the first observation of this effect.
}\label{fig:Data_Scan}
\end{figure}

A run begins with the lasers locked to \PBC\ modes that are separated by $4\times c/2L=\nm{2}{0}{GHz}$ (\figref{fig:Tuning_Scheme}, left).  The cavity and lasers are then swept over L1, acquiring a calibration spectrum with $\delta\ts{cal}=2\pi\times1\text{ GHz}$. The ``\charybdis" laser is moved to the same longitudinal \PBC\ mode as the ``\scylla" laser. Then, both lasers are tuned back-and-forth over the limit point L2 many times, acquiring data with  $\delta\ts{lim}\le2\pi\times2\text{ MHz}$.  During the $\approx$1 hour that a run can take, the lasers' intensities in the \PBC\ may droop a few percent due to mechanical drift and dye consumption.  The \PBC\ transmission signals of both lasers are continuously recorded and used to correct for the droop.  \scylla's transmission peaks through a stabilized reference cavity are recorded and used to calibrate the frequency axis and combine the data into a single spectrum, such as shown in \figref{fig:Data_Scan}.  The stability of the atomic-beam density is monitored by measuring absorption on the $6s^2$ $\ssz\to6s6p$ $\spo$ transition and found to vary negligibly.
%
%
%
%
%
%

To extract an estimate of $\nu$ from a run, a model lineshape $\scL(\Omega_1)$ is constructed from the calibration spectrum.  The data in the limit region are then fit to the model function $a\,\scL\prn{\Omega_1-\Omega_{31}/2}+b\,\Omega_1+c$, where the parameters $a$, $b$, and $c$ are determined by the fit, and $\Omega_{31}/2$ is the frequency difference between the calibration peak at L1 and the limit point at L2.  Without detailed knowledge of the light shifts, the exact frequency of the calibration peak and the limit point are uncertain.  But all that is required is to know the frequency difference between them, and a simple measurement determines this:  Tuning the lasers from L1 to L3 (\figref{fig:Tuning_Scheme}), and continuously acquiring Scylla's reference-cavity peaks, two identical calibration spectra are acquired, one at L1 and another at L3, separated by $\Omega_{31}\cong41.5$ reference-cavity peaks.  Half this value is the frequency difference between the calibration peak and the limit point.

\begin{figure}
\includegraphics[width=240pt]{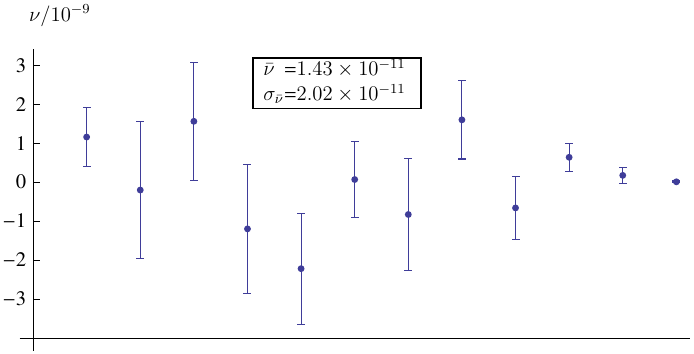}
\caption{(color online). $\nu$ from twelve runs.  The error bars decrease over the lifetime of the experiment as improvements in excitation and detection efficiency were achieved.}\label{fig:NuPlot}
\end{figure}

The best-fit peak amplitude $a$ is the ratio ${\scS\ts{lim}}/{\scS\ts{cal}}$. Each run produces a best-fit estimate of $\nu$. The values of $\nu$ from twelve runs, the weighted average, and its standard error, are shown in \figref{fig:NuPlot}. The best-fit value of $\nu$ is $\prn{1.4\pm2.0}\times10^{-11}$, limiting the relative rate to $\nu<\nm{4.0}{-11}{}$ at the 90\% confidence level.
%
%
%
%
%
%
%
%
%

A number of sources of systematic error must be considered.  The Landau-Yang theorem is circumvented in the nonzero-nuclear-spin isotopes of barium, $\tu{137}$Ba (11\% natural abundance, $I=3/2$) and $\tu{135}$Ba, (6.6\%, $I=3/2$),  by two manifestations of the hyperfine-interaction (\HFI):  \HFI-induced splitting of the intermediate $6s6p$ \spo\ state, and \HFI-induced mixing of states with $J'=2$ (primarily $6s7d$ \specterm3D2) with the final $5d6d$ \specterm3D1 state \cite{KOZ09}.  The amplitudes for both processes are comparable, and the rate, relative to an allowed transition (i.e., expressed in the same way as the limit parameter $\nu$), is $\sim10^{-8}$.
This rate is $10^3$ times larger than the present level of sensitivity, and the \HFI-induced transitions are observed (e.g., in the right side of \figref{fig:Data_Scan}). To our knowledge, this is the first observation of \HFI-enabled two-photon transitions.  Fortunately, the peaks of the \HFI-induced transitions, although close to the limit region, do not completely obscure it. As is also discussed in Ref. \cite{KOZ09}, magnetic fields can create a false-positive signal in a way similar to the \HFI.  A relatively large field ($\sim$10 gauss) would be necessary to generate a  false-positive signal at the present level of sensitivity.

%
%
%
Another systematic effect, the one that limited the sensitivity in Ref. \cite{DeM99}, is the non-zero spectral width of the lasers.  Photons in opposite wings of the lasers' spectral profiles can together drive the nominally \BE-statistics-forbidden transition at a relative rate $\sim(\Gamma\ts{L}/\Delta)^2\sim\nm{1}{-12}{}$, where the laser line width is $\Gamma\ts{L}\approx2\pi\times3$ MHz.
%
%

When two counter-propagating light beams drive a two-photon transition, the Doppler effect due to the motion of the atoms collinear with the light (i.e., the transverse motion in the atomic beam) shifts the frequencies, as seen by the atoms, of the two light fields in opposite directions. This leads to a nonzero allowed two-photon-transition amplitude. It is, however, largely suppressed in the case of a power-buildup cavity \cite{Eng07}.  The residual effect has a relative rate  $T^2\,{\Gamma\ts{D}^2}/{\Delta^2}\sim\nm{2}{-15}{},$ where $T$=\nm{1}{-2}{} is the \PBC\ mirror transmission, and $\Gamma\ts{D}$=2$\pi\times$13 MHz is the Doppler width of the two-photon transition.

%
%

At the limit point, the amplitudes that destructively interfere ($\scA_{jk}$ in \figref{fig:transition}) proceed through  magnetic sublevels (or orthogonal superpositions thereof) of the intermediate state.  A disturbance of the degeneracy of these sublevels may alter the balance of the two amplitudes, preventing perfect cancelation, and generate a false-positive signal in a way similar to the \HFI-splitting mentioned above.  
Considering only the states in \figref{fig:transition}, light shifts produce a difference in the intermediate state energies of $\delta_i=-\prn{\sD_{ng}^2-\sD_{en}^2}\prn{E_1^2-E_2^2}/4\hbar\Delta$, where $E_1$ and $E_2$ are the electric field intensities of the light.  Notable is that both an imbalance in the light intensities and an imbalance in the dipole moments are necessary to make $\delta_i\neq0$.  There are, besides, additional atomic states in the region around \ket{e} that make smaller but significant contributions to $\delta_i$.  %
Roughly calculated, %
the relative-rate due to light shifts is $(\delta_i/\Delta)^2\sim10^{-12}$, an order of magnitude below the present level of sensitivity.

%
%
The analysis of \eqnref{eq:twophotrate} and sequelae considered only E1E1 two-photon transitions.  For the geometry of this experiment, where the light beams are collinear, the analysis still holds true for all multipole combinations \cite{Eng07}.  However, misalignment of the laser beams permits certain higher-order multipole transitions \cite{Eng07}.  The leading non-zero terms, E1M2 and E2M1 \cite{Dun04}, roughly of the same magnitude, contribute a false-positive signal at a rate $\sim\abs{\sD_{M2}/\sD_{E1}}^2\sin^4\theta$ $\sim10^{-10}\theta^4\sim10^{-19}$, where $\sD_{M2}$ and $\sD_{E1}$ are the magnetic-quadrupole and electric-dipole reduced matrix elements of the transition, and $\theta\approx0.3^\circ$ is a conservative upper bound on the misalignment angle.

In conclusion, the reported experiment has improved the limit on possible Bose-Einstein statistics violation by photons by more than three orders of magnitude.  Additionally, we have observed hyperfine-interaction induced two-photon transitions.  In principle, further improvement by an order of magnitude or more is possible with this technique by improving laser-frequency lock which will allow longer statistics accumulation.

We are grateful to D. DeMille, M. Auzinsh, M. Kozlov, and M. Zolotorev for inspiration and support, and to undergraduates L. Zimmerman, Y. Rosen, and K. Choi for hands-on contributions. This research was supported by NSF.
\bibliography{Spectroscopic_test_of_Bose_English_Budker}
\end{document}